\documentclass[apj]{emulateapj}

\newcommand{\mpch}{\>h^{-1}{\rm {Mpc}}}

\def\gtsima{$\; \buildrel > \over \sim \;$}
\def\ltsima{$\; \buildrel < \over \sim \;$}
\def\prosima{$\; \buildrel \propto \over \sim \;$}
\def\gsim{\lower.7ex\hbox{\gtsima}}
\def\lsim{\lower.7ex\hbox{\ltsima}}
\def\simgt{\lower.7ex\hbox{\gtsima}}
\def\simlt{\lower.7ex\hbox{\ltsima}}
\def\simpr{\lower.7ex\hbox{\prosima}}

\usepackage{graphicx}

\newcommand{\be}{\begin{equation}}
\newcommand{\ee}{\end{equation}}

\def\apj{ApJ}

\def\aj{AJ}
\edef\aap{A\&A}
\def\mnras{MNRAS}

\begin{document}  
 
\shorttitle{Detection  of the Acoustic Peak}
\shortauthors{Mart\'{\i}nez et al.}

\title{Reliability of the detection of the Baryon Acoustic Peak }

\author{Vicent J. Mart\'{\i}nez\altaffilmark{1,2},
Pablo Arnalte-Mur\altaffilmark{1,2},  
Enn Saar\altaffilmark{3}, 
Pablo de la Cruz\altaffilmark{1},
Mar\'{\i}a Jes\'us Pons-Border\'{\i}a\altaffilmark{4}, 
Silvestre Paredes\altaffilmark{5},
Alberto Fern\'andez-Soto\altaffilmark{6},
and
Elmo Tempel\altaffilmark{3}
}

\altaffiltext{1}{Observatori Astron\`omic, Universitat de Val\`encia, Apartat
de Correus 22085, E-46071 Val\`encia, Spain}
\altaffiltext{2}{Departament d'Astronomia i Astrof\'{\i}sica, Universitat de Va\-l\`en\-cia, 46100-Burjassot, Val\`encia, Spain} 
\altaffiltext{3}{Tartu Observatory, EE-61602 T\~oravere, Estonia}
\altaffiltext{4}{Secci\'on Departamental de Matem\'atica Aplicada,
Escuela Universitaria de Estad\'{\i}stica, 
Universidad Complutense de Madrid, Avda. Puerta de Hierro s/n, 
28040 Madrid, Spain}
\altaffiltext{5}{Departamento de Matem\'atica Aplicada y Estad\'{\i}stica, 
Universidad Polit\'ecnica de Cartagena, C/ Dr. Fleming s/n,
30203 Cartagena, Spain}
\altaffiltext{6}{Instituto de F\'{\i}sica de Cantabria (CSIC-UC)
Edificio Juan Jord\'a,  Av. de los Castros s/n, E-39005, Santander, Spain}

\begin{abstract} 
The correlation function of the distribution of matter in the
universe shows, at large scales, baryon acoustic oscillations, which
were imprinted prior to recombination. This feature was first detected
in the correlation function of the luminous red galaxies (LRG) of the Sloan
Digital Sky Survey (SDSS). The final release (DR7) of the SDSS has
been recently made available, and the useful volume is about two times
bigger than in the old sample. We present here, for the first time, the 
redshift space correlation function of this sample at large scales
together with that for  one shallower, but denser volume-limited
subsample drawn from the 2dF redshift survey. We test the reliability
of the detection of the acoustic peak at about $100 \, \mpch$ and the
behaviour of the correlation function at larger scales by means of careful estimation
of errors. We confirm the presence of the peak in the latest data although broader than in previous detections. 
\end{abstract}


\section{Introduction}
\label{sec:intro}
About 380,000 years after the Big Bang, when the temperature falls low
enough for recombination to occur, the matter in the Universe becomes
neutral. At this epoch, the sound speed drops off abruptly and
acoustic oscillations in the fluid become frozen. Their signature can
be detected in both the Cosmic Microwave Background (CMB) radiation
and the large-scale distribution of galaxies.  The characteristic
length scale of the acoustic oscillations can be used as a ``standard
ruler" to probe the expansion history of the universe.

The imprints in the matter distribution of this feature, called
baryon acoustic oscillations (BAO), should be detectable in both the
correlation function and the matter power spectrum. Moreover, this
feature should manifest itself as a single peak in the correlation
function at about $100 \, \mpch$\footnote{$h$ is the Hubble constant in 
units of 100 km s$^{-1}$ Mpc$^{-1}$.}. The first unambiguous detection of
this feature in the correlation function (redshift space) of the Sloan Digital Sky
Survey Luminous Red Galaxies (SDSS-LRG) was reported by
\citet{2005ApJ...633..560E}. Detection of these oscillations in the
power spectrum of the galaxy distribution was first reported by
\citet{2005MNRAS.362..505C} for the 2-degree Field Galaxy Redshift
Survey (2dFGRS). Cosmological parameters can be determined from
the position of the baryon acoustic 
peak \citep[see, e.g.,][]{2007ApJ...657...51P}.

In this letter we study the correlation functions of the largest
available redshift surveys (2dFGRS and SDSS).  We focus on the
two-point correlation function, and study the reliability of the
detection of the BAO feature at $100 \, \mpch$ in the samples drawn
from those surveys. In our calculations, we
adopted a cosmological model with $\Omega_M = 0.27$, and
$\Omega_{\Lambda} = 0.73$.

\section{Data}

In the first detection of the baryon acoustic peak,
\citet{2005ApJ...633..560E} used a sample that was constructed
selecting about 15 luminous red (early-type) galaxies per square
degree, using different luminosity cuts
\citep[see][]{2001AJ....122.2267E}.  For correlation studies, an
almost volume-limited (constant-density) subsample was chosen, its
characteristics are listed in Table~\ref{tab:data} as DR3-LRG.

\label{sec:data}
\begin{table*}
\begin{center}
\caption{\sc Characteristics of the samples used and quoted\label{tab:data}}
\begin{scriptsize}
\begin{tabular}{lrcccll}
\tableline\tableline
Sample & Number& Magnitude limits$^a$& Redshift limits& Solid& Volume&number
density\\
&of galaxies & & &angle (sr)& ($h^{-3}$Gpc$^3$)& ($h^{3}$ Mpc$^{-3}$) \\
\tableline
\\
DR3-LRG & 46,748 & $-23.2 < M_g(z=0.3) < -21.2$ &$0.16 < z < 0.47 $&1.16&0.75 $^b$&$6.3\times 10^{-5}$ $^b$ \\
DR7-LRG & 102,568 & $-22.344 < M_g(z=0.3) < -20.344$ &$0.16 < z < 0.47 $&2.19& 1.41 &$7.3 \times 10^{-5}$ \\
DR7-LRG-VL & 44,164 & $-23.544 < M_g(z=0.3) < -21.644$ &$0.14 < z < 0.42 $&2.19&1.03&$4.3\times 10^{-5}$ \\
2dFVL & 33,878 & $-21 < M_{b_J} < -20$ &$0.03< z < 0.19$&0.45& 0.023 & $1.5 \times 10^{-3}$ \\
\\
\tableline
\end{tabular}
\end{scriptsize}
\tablecomments{$^a$Absolute magnitudes $M$ are normalized to $H_0=100$ km s$^{-1}$ Mpc$^{-1}$. $^b$ Using $\Omega_M = 0.3$, $\Omega_{\Lambda} = 0.7$ as in
\citet{2005ApJ...633..560E}, these values are $0.72$ and $6.5\times
10^{-5}$, respectively.}
 
\end{center}
\end{table*}

The last data release \citep[DR7, see][]{2008arXiv0812.0649A} of the
SDSS-LRG contains spectra for 206,797 LRGs within a solid angle of
9,380 square degrees. We select a subsample of these data (the main
compact body of the sky footprint, 7,204 square degrees) in order to
minimize the influence of border corrections on our results.  Choosing
the same redshift and magnitude limits as \citet{2005ApJ...633..560E} 
and slightly shifted magnitude limits (see Table 1) we have a
sample that is about twice as large as the one
used by them; we label it as DR7-LRG in Table~\ref{tab:data}. This
sample is approximately volume-limited, but not in a formal way.
We have determined a real volume-limited (VL) 
sample \citep[as in][]{2005ApJ...621...22Z}, with nearly constant comoving 
number density,  applying the appropriate luminosity cut
and restricting the distance range to ensure completeness;
it is labelled as DR7-LRG-VL. 

For comparison, we use a nearly volume-limited sample from the
full 2dFGRS prepared by the 2dF team \citep{2004MNRAS.352..828C}. It
contains luminous galaxies in two spatial slices. The number density 
in the 2dFVL sample is more
than an order of magnitude larger than the number density of the
SDSS-LRG samples used here (see Table~\ref{tab:data}).

Although the volume covered by the 2dFVL sample is smaller than that 
of the DR7-LRG, it is still useful to measure correlation on 
scales $\sim100\,\mpch$. As shown below, cosmic variance is not dominant, even if
its effect is an order of magnitude larger for 2dFVL than for DR7-LRG. Moreover, the 
increase in density compensates the decrease in volume, so that the number of
pairs of galaxies in each distance bin at these scales is similar in both samples. 
Hence, discreteness errors should be similar in both cases. 

\begin{figure*}
\centering
\resizebox{0.9\textwidth}{!}{\includegraphics*{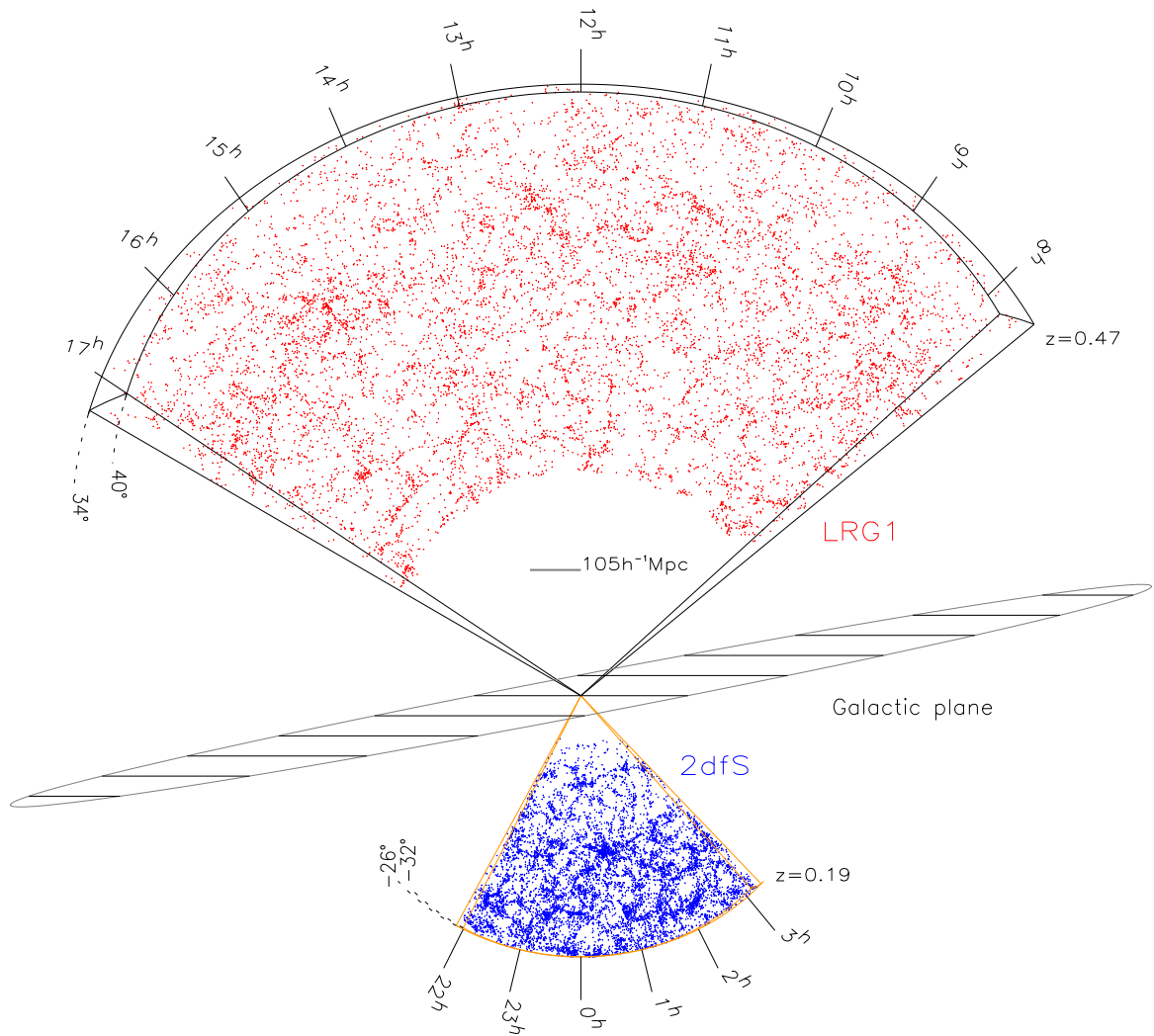}}
\caption{
The large slice is drawn from the SDSS-LRG (DR7) survey. The slice is $6^\circ$ wide in declination and the galaxy distribution is shown within the redshift range $0.16\leq z \leq0.47$. There are 10,136  red luminous galaxies within this slice depicted as red dots.
The smaller slice with blue dots shows the galaxy distribution of 9,744 objects from the Southern Galactic hemisphere of the 2dFVL sample, reaching a depth of $z=0.19$.
To illustrate the scale of the acoustic peak a segment of length $105\,h^{-1}$Mpc is shown to scale.}
\label{fig:slices}
\end{figure*}

These three samples are important for the detection of the acoustic
peak. Although the SDSS-LRG samples are larger and cover a redshift range less
affected by nonlinear effects, the lower redshifts mapped by the 2dFVL
sample are also important, providing the yardstick that can be compared with the
characteristic BAO scales at larger redshifts (see Fig.~\ref{fig:slices}).

\section{Estimating correlation functions}

We estimated the spherically-averaged
redshift-space correlation functions by using the Landy-Szalay
border-corrected estimator \citep{1993ApJ...412...64L} that has good
large-scale properties.  We generated a random distribution of points
following the selection function of each catalogue considered, and
estimated the correlation function $\xi(s)$. 
\begin{equation}
\widehat{\xi}_{\rm{LS}} (s) = 1 +
\frac{DD(s)}{RR(s)} - 2\frac{DR(s)}{RR(s)} \,
\label{eq:ls}
\end{equation}
where $DD(s)$, $RR(s)$ and $DR(s)$ are the probability densities of galaxy-galaxy,
random-random and galaxy-random pairs, respectively, for a pair distance 
$s$. There are several recipes for the choice of the size $N_{\mathrm{rd}}$ of 
the random point set; as we are interested in large-distance correlations, where 
the numbers of pairs per bin (kernel width) are large, we 
used $N_{\rm{rd}}\approx 5 N$ ($N$ is the number of galaxies in the sample). 
Increasing $N_{\rm{rd}}$ up to $20N$ led to pointwise differences less than a percent.

We estimate the probability densities by the kernel method, summing the
box spline $B_3(\cdot)$ kernels \citep{2007MNRAS.374.1030S} centered at 
each pair distance, and sampling the distributions at smaller intervals than the kernel width.

For the 2dFVL sample, we generated the random catalogues using the
subroutines to calculate the completeness and the magnitude limit from
the angular masks, both provided by the 2dF team
\citep{2001MNRAS.328.1039C}.

For the DR7-LRG samples we generated the angular mask from the
data \citep[as in][]{2006A&A...449..891H}, following the scan stripes of the survey and defining by hand
the rectangles in the survey coordinate system $\eta,\lambda$ that
cover the data footprint. We assumed the angular selection to be
constant inside this mask.

The DR7-LRG sample is only approximately vo\-lume-lim\-ited, and we
generated the random samples based on the smoothed comoving density of
the data.
The accepted statistical paradigm for the galaxy distribution is to model
it as a Cox process -- a Poisson point process where the local intensity is
determined by a realization of a random field. The two-stage nature of
this process leads to two sources of errors of sample statistics (correlation
functions, in our case). The first source is the possible deviation of the realization correlation function from the true one (cosmic variance) and the other source is
the discreteness of the point process -- how well its correlation function
estimate approximates that of the particular realization. 

Precise estimates of the cosmological 
statistics have become actual;  \citet{2008arXiv0810.1885N} classifies the error estimates as internal and external,
this coincides with our classification. As cosmic variance and discreteness variance are independent, they add for the total variance.

If the random field is Gaussian, as usually assumed, and its expected correlation function and power spectrum are more or less known, the covariance of the correlation function estimate can be found as a convolution of the correlation function itself, or
as an integral over the power spectrum squared  \citep[see, e.g.,][]{2006NewA...11..226C}.
This is common for all samples; a rough upper limit is
\begin{eqnarray*}
\mathrm{Var}(\widehat\xi(r))&<&\frac1{2\pi^2}\frac1{Vr^2}\int P^2(k)\,dk\approx\\
&\approx& 5\cdot10^{-8}(V/h^{-3}\,\mathrm{Gpc}^3)^{-1}(r/100\,h^{-1}\,\mathrm{Mpc})^{-2},
\end{eqnarray*}
where the numerical value was obtained by using the \citet{1998ApJ...496..605E} approximation for the power spectrum, with a Gaussian cutoff at $k=100\,h\,\mathrm{Mpc}^{-1}$ (the real-space scale of about $60\,h^{-1}\mathrm{Mpc}$). The sample volume $V$ and pair distance $r$ in this formula are typical for the DR7-LRG and the baryonic peak; the rms error is about $2\cdot10^{-4}$ in this case, and about $1.5\cdot10^{-3}$ for the 2dFVL sample. As we shall see below, this is much smaller than the discreteness error in both cases, and we shall neglect it for the rest of the paper.

The discreteness error can be estimated in several ways. The most
attractive of these is bootstrap, that uses only the knowledge inherent
in the observed data; the observed sample is repeatedly resampled and
the statistics found averaging over these bootstrap samples. The assumption that the observed values are i.i.d. (independent and identically distributed) demands that resampling is done with replacement; this leads to the fact that about 1/3 
of sample data is sampled more than once.
Bootstrapping has been applied to estimate correlation function errors in
cosmology before, first by \citet{1984MNRAS.210P..19B}, but it is intuitively clear that the bootstrap samples represent a different world, where 1/3 of the galaxies of the original sample coincide (are close pairs). So it has been avoided lately, and other methods (block jackknife, for example) are being used. 

The latter is a step in the right direction. The reason 
why the direct approach to bootstrap correlations fails is that bootstrap estimates
can be applied only to smooth functions of sample means \citep[see,
e.g.][]{tibshirani, lahiri}. Only in this case can the estimates be
proved to be consistent (approaching the population statistics in the
large sample limit). Correlation functions do not belong to this
class.

However, there is an elegant way to solve the problem \citep{lahiri}. Let us take a simple case of a random process $Y(x_i)$ defined on an onedimensional grid. Then, for studying its correlation function at the lag $k$, create another, two-dimensional process $Z_2(x_i)=(Y(x_i),Y(x_{i+k}))$. Averaging over the product of the components of $Z_2$ gives us the covariance function.

Generalization  of this approach to 3-D point processes  involves introducing a fine 3-D grid, and 
generating a random process $Y(x_i)$ on the grid by 
assigning to its vertices values 0 or 1 depending if the grid point (or cell) hosts a point of our point process.

As above, let us create a new multidimensional process $Z_m(x_i)$, a collection of
$m$ values of $Y(\cdot)$ at the grid points at a fixed distance interval (bin) from $x_i$.
The number of points in the bin, if the original cell at $x_i$ contains a point (the number of neighbours) 
can be defined as a simple function of this process, $N(Z_m(x_i))$.
The sample mean of this function is the mean number of neighbours for a given distance
bin, proportional to the function $DD(s)$ in Eq. (\ref{eq:ls}).
This function can be bootstrapped. Notice that if we consider all possible bins we get a full histogram of neighbours of a sample point. 
Thus, the natural things to
bootstrap are such pointwise histograms -- histograms of all the neighbour distances from a given point, and, if needed, histograms of distances of random points from a given sample point. All points carry their histograms; generating a bootstrap realization of a correlation function reduces to selecting a sample of histograms (with replacement), summing them to get the bootstrap realization, and applying, e.g., Eq. (\ref{eq:ls}) to get the bootstrap realization of a sample correlation function. 
In order to be comparable to the original data, the number of histograms in the
sum have to be the same as the number of points in the sample; as usual, about 1/3 of them occur in the sum more than once. Random point pairs should not be bootstrapped, as these are merely a tool for performing volume integrals.

This approach is logical and easy to apply. However, not even these histograms
are independent, as required -- the locations of sample points are correlated,
and this has to be taken into account. The recipe for that is not to bootstrap 
individual points, but data blocks \citep{lahiri}. Such a procedure can be proved to give consistent estimates. The size of the block is determined by the correlation length. We shall show below that it is the case.

An unusual (for cosmologists) feature of block bootstrap is that blocks can
overlap; in fact, the estimates with overlapping blocks are more efficient than
those with non-overlapping blocks. 

Numerically, this procedure is much cheaper to apply than that of
\citet{1984MNRAS.210P..19B}. If we have enough core memory, we calculate 
and store the pair distance probability densities for every sample point, 
average them for block densities, and sum these for bootstrap realizations. 
As we do not need to recalculate distances, this bootstrap algorithm is fast.
It is straightforward to modify this procedure for pairwise or pointwise weighting.

\section{Application to Poisson-Voronoi processes}

We demonstrate and check the bootstrap machinery described above on a
point field where the correlation function is known exactly
-- the Poisson-Voronoi process. As this process does not depend on the
underlying random field, this is a clean and direct check of the
recovery of the discreteness variance.

Following \citet{1989A&A...213....1V}, we start from a set of seed
points distributed randomly in space (a Poisson point 
process). The set of all vertices of their Voronoi tessellation
defines another (Poisson-Voronoi vertices) point process with a well known numerically tractable
two-point correlation function \citep{heinrich98}. 

We used this process as a testbed for our correlation function
estimates and the bootstrap recipes for its variance and bias. For
that, we generated 100 samples of Voronoi vertices within the
DR7-LRG sample volume, with about the same mean density. We found
correlation functions for all these realizations, estimated their bias
and variance, and found the symmetric confidence regions.  

\begin{figure}
\centering
\resizebox{0.5\textwidth}{!}{\includegraphics*{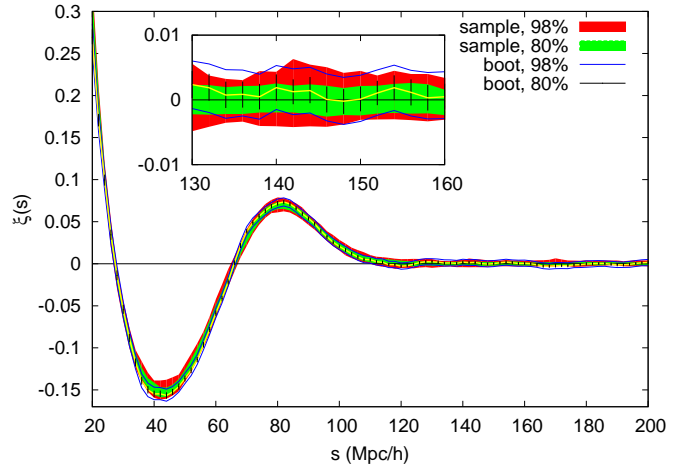}}
\caption{Comparison of the sample variability and bootstrap
predictions for a Poisson-Voronoi process modelling the galaxy
distribution in the DR7-LRG sample volume. The variability of a sample
of 100 realizations of the process is shown by color, the 98\% confidence
region by red and the 80\% confidence region by green. The block bootstrap
estimates of variance are made on the basis of a single realization
(its correlation function is shown by the yellow line), the predicted
confidence regions are shown by blue lines (98\% confidence) and by
errorbars (80\% confidence). The figure shows the overall fit of the
observed and predicted variability, the inset shows that for a smaller
region at a larger scale.}
\label{fig:corrvor}
\end{figure}

Fig.~\ref{fig:corrvor} shows the 98\% and 80\% confidence regions
obtained from these realizations. Then we selected one of these realizations and
looked how well the bootstrap procedure predicts the real bias and
variance. The predicted variance was the closest to the true one for
the block radius $R=15\,h^{-1}$Mpc, which practically coincides with
the correlation length ($\xi(R)\approx 1$) for our process. The
bootstrap confidence limits are shown in Fig.~\ref{fig:corrvor} by the
blue lines (for 98\%) and by errorbars (for 80\%); the variances are
easier to compare in the inset. Biases were always considerably
smaller than variances, both for the true samples and for the
bootstrap samples. Based on these tests, we feel that the bootstrap procedure
works well; we shall apply it to the galaxy correlation
functions below.

\section{Large-scale galaxy correlation functions}

\begin{figure}
\centering
\resizebox{0.5\textwidth}{!}{\includegraphics*{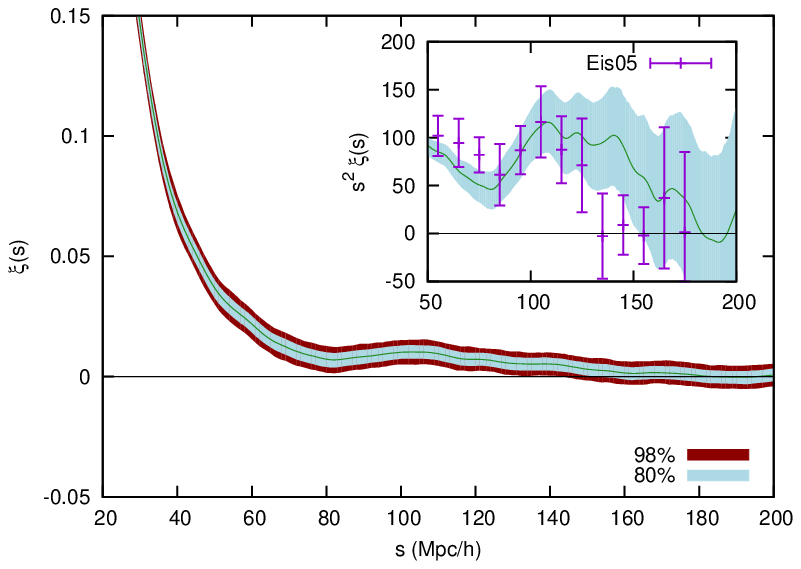}}
\resizebox{0.5\textwidth}{!}{\includegraphics*{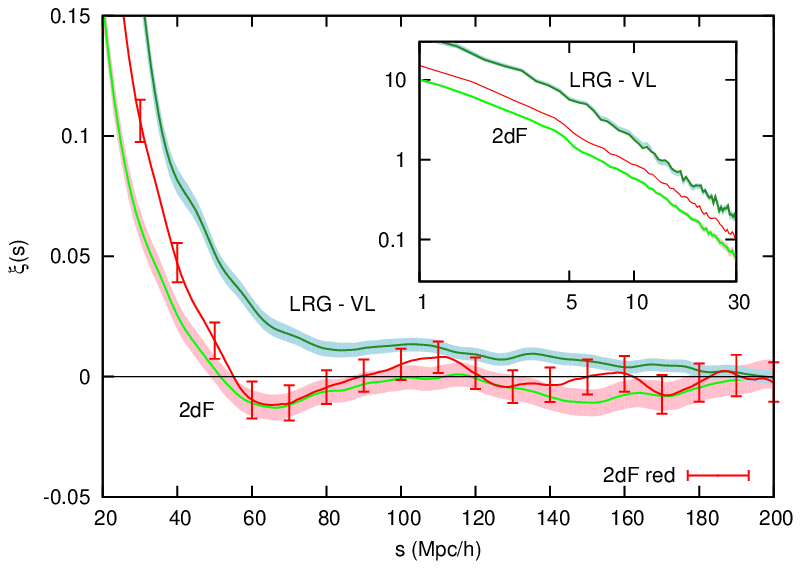}}
\caption{The redshift space correlation function for the different
samples.  The top main panel shows the correlation function for the DR7-LRG
sample (the 98\% and 80\% confidence regions in dark red and pale blue and the
function as a green line). The acoustic bump is clearly seen at about $105 \, \mpch$.
The inset shows the large-scale behaviour of the correlation function
(multiplied by $s^2$). The function is shown by the green line, its
80\% confidence region in pale blue, and it is compared with the original
BAO discovery data \citep{2005ApJ...633..560E} with errorbars, labelled by Eis05. Note that 
the the acoustic peak is wider in the new sample. The bottom panel shows the redshift-space correlation function for the volume-limited samples: for the 2dFVL sample (the 80\% confidence region
in pink, the function in light green), for the red galaxy subsample of
the 2dFVL (the red line with errorbars covering the 80\% confidence
region), and for the DR7-LRG-VL (the 80\% confidence region
in pale blue, the function in dark green). The inset shows these 
correlation functions at short scales. Numerical values of the correlation functions
plotted here can be downloaded from \texttt{http://www.uv.es/martinez}.}
\label{fig:corrgal1}
\end{figure}

The redshift-space correlation functions for our samples are
summarized in Fig.~\ref{fig:corrgal1}. 
The correlation functions are
found as described above, and their errors (biases, variances,
confidence regions) are estimated by block bootstrap, with blocks as
spheres with the radius $R=12.0\,h^{-1}$ Mpc for the DR7-LRG sample
and $R=6.5\,h^{-1}$ Mpc for the 2dFVL sample (the respective
correlation lengths).  We show in the bottom panel also the correlation function for
a subsample of red galaxies from the 2dFVL \citep[chosen according to 
the spectral type of the galaxy, see][]{2002MNRAS.333..133M}, with 17,252
galaxies. The block size was chosen the same as for the main sample.

The top panel shows the correlation function for the DR7-LRG sample. We see that
the function $\xi(s)$ remains
positive in the whole range shown (up to $200\, \mpch$). It also
shows confidently the $105$ $\mpch$ maximum.
The inset amplifies the large-distance behaviour of the LRG
correlation function.  We show the quantity $s^2\xi(s)$ for the DR7-LRG 
and compare it with the correlation function estimates by
\citet{2005ApJ...633..560E}. We show here only the 80\% symmetric
confidence region that is comparable with the $\pm\sigma$ limits of
the previous estimate. We see that the new data shows the BAO peak
with a much better confidence and also reveals that the peak is much wider than the 
peak found in DR3-LRG. A similar trend was also seen in the analysis
by \citet{2009MNRAS.393.1183C}.

The bottom panel shows the 80\% confidence level region for the 2dFVL (by pink),  
and shows the 80\% confidence limits
for the red galaxies from the 2dFVL by errorbars.
In contrast to the DR7-LRG, the correlation function for the
2dFVL sample crosses zero at about $55 \, \mpch$ reaching a local minimum
with $\xi(s) <0$ at about $65 \, \mpch$. At larger scales, this
correlation function tends also towards the $105 \, \mpch$ maximum. This
maximum is present, even more prominently, in the correlation function
of the red galaxy
subsample. It is precisely at this scale where the LRG sample shows
the acoustic baryonic peak first noted by \citet{2005ApJ...633..560E}.

The inset shows the behaviour of the correlation function of the
volume-limited samples, DR7-LRG VL, 2dFVL, and 2dFVL red, 
at short scales,  characterised by 
a power-law regime with a downturn at the smallest 
scales \citep[see, e.g.,][]{2007MNRAS.381..573R}. The
difference in the correlation function amplitudes is a clear
fingerprint of the luminosity-color segregation. The LRG galaxies
are very luminous red objects lying in more dense environments and
displaying enhanced clustering.

\section{Conclusions}
\label{sec:dis}
The results show, first, that the baryon acoustic peak is a stable feature in
the large-distance galaxy correlation function. We have presented 
the analysis of the LRG sample drawn from the last data release
(DR7) of the SDSS. This survey 
is much larger and more compact than the data used before, 
so edge-correction effects are smaller now and results are therefore 
more robust. While the peak is most prominent in the distribution of luminous red galaxies 
(because of the large spatial extent of the sample), it can be seen, if we know what to look for, in
smaller galaxy samples. We have reported the first detection of the baryon acoustic peak  
in a volume-limited sample of very luminous galaxies drawn from the 2dFGRS. 

Second, the baryon peak is much wider than found before and than
expected; the distortions responsible for that demand careful analysis.

Third, the minimum in the large-distance correlation
functions of some samples demands explanation -- is it
really the signature of voids?

And, fourth, a new internal method to estimate the errors of the 
correlation function has been introduced. The method is based
on a generalization of the application of the bootstrap resampling
techniques to smooth functions of sample means.

\smallskip
\textbf{Acknowledgements} 

We thank Darren Croton for the 2dfGRS samples and the mask data,
Rien van de Weygaert and Dietrich Stoyan for their advise regarding 
the Voronoi-Poisson process, Robert Lupton for useful discussions and 
Jun Pan and Jan Hamann for insightful comments on a preliminary draft of this letter.

This work has been supported by the University of Valencia 
through a visiting professorship for Enn Saar and by the Spanish MEC projects
AYA2006-14056 and CSD2007-00060, including
FEDER contributions. PAM acknowledges support from the
Spanish MEC through a FPU grant. ES and ET acknowledge support by
the Estonian Science 
Foundation, grants No. 6104, 6106, 7146 and by the Estonian Ministry
for Education and Science, grant SF0060067s08.

Funding for the SDSS and SDSS-II has been
provided by the Alfred P. Sloan Foundation, the Participating
Institutions, the National Science Foundation, the U.S. Department of
Energy, the National Aeronautics and Space Administration, the
Japanese Monbukagakusho, the Max Planck Society, and the Higher
Education Funding Council for England. The SDSS Web Site is
\texttt{http://www.sdss.org/}.


\end{document}